\documentclass[debug]{rmaa}

\usepackage{paralist}

\usepackage{psfrag,color}

\usepackage[latin1]{inputenc}

\title{Compact Radio Sources in the Field of Tycho's Supernova Remnant} 

\author{
  Luis F. Rodr\' \i guez,\altaffilmark{1} 
  Vanessa Yanza,\altaffilmark{1}
  and Sergio A. Dzib\altaffilmark{2}}

\altaffiltext{1}{Instituto de Radioastronom\'\i{}a y Astrof\'\i{}sica,
  UNAM, M\'exico.}

\altaffiltext{2}{Max-Planck-Institut f\"{u}r Radioastronomie, Germany.}

\shortauthor{Rodr\' \i guez, Yanza, \& Dzib}
\shorttitle{Compact Radio Sources in the Field of Tycho's SNR}

\fulladdresses{
\item Sergio A. Dzib: Max-Planck-Institut f\"{u}r Radioastronomie, Auf dem H\"{u}gel 69, D-53121 Bonn, Germany.
\item Luis F. Rodr\' \i guez and Vanessa Yanza: Instituto de Radioastronom\'\i{}a y
  Astrof\'\i{}sica, Universidad Nacional Aut\'onoma de M\'exico,
  Apartado Postal 3--72, 58090 Morelia, Michoac\'an, M\'exico
  (l.rodriguez@irya.unam.mx).}

\listofauthors{Luis F. Rodr\' \i guez, Vanessa Yanza, \& Sergio A. Dzib}
\indexauthor{Rodr\' \i guez, L. F.}
\indexauthor{Yanza, V.}
\indexauthor{Dzib, S. A.}

\abstract{
We present sensitive, high angular resolution Jansky Very Large Array 
observations made in 2014 at 1.50 GHz 
toward the field of Tycho's supernova remnant. We detect a total of 36 compact sources in a field
with radius of 13 arcmin. This number is consistent with the expected number of background sources. We use 
older observations made with the classic Very Large Array to compare with the 2014
observations and search for sources showing large proper motions that could be related to 
the donor companion of the exploding white dwarf that produced the supernova in 1572.
The comparison of the positions for the two sets of observations
does not show sources with large proper motions and supports the conclusion that all sources detected are extragalactic
and unrelated to the supernova field.}

\resumen{
Presentamos observaciones sensitivas de alta resoluci\'on angular hechas en 2014 con el Jansky Very Large Array
 a 1.50 GHz hacia el campo de la remanente de supernova de Tycho. Detectamos un total de 36 fuentes compactas en un
 campo con radio de 13 minutos de arco. Usamos observaciones mas antiguas hechas con el VLA cl\'asico 
 para comparar con los datos de 2014 buscando
 fuentes con movimientos propios grandes que pudieran estar relacionadas con la compa\~nera donante de la enana blanca que
 explot\'o y produjo la supernova en 1572. La comparaci\'on de posiciones entre los dos conjuntos de observaciones
 no revela fuentes con grandes movimientos propios y apoya la conclusi\'on de que todas las fuentes detectadas
 son fuentes extragal\'acticas de fondo sin relaci\'on con el campo de la supernova.}

\addkeyword{ISM: supernova remnants}
\addkeyword{stars: general}
\addkeyword{astrometry}
\addkeyword{radio continuum: general}

\begin{document}
\maketitle

\section{Introduction}
\label{sec:intro}

Tycho's 1572 supernova (SN 1572, 3C 10, B Cas) has been classified as a standard type Ia on the basis of its light-echo spectrum 
(Krause et al. 2008). This supernova type is very important since it is a cosmological standard candle that has allowed to determine the accelerated expansion history of the Universe (Riess et al. 1998; Perlmutter et al. 1999). It is produced
by the thermonuclear explosion of a white dwarf accreting from a companion in a close binary system. In the most favored model the white dwarf is fully disrupted when it reaches the Chandrasekhar limit, leaving no collapsed object. The companion
then leaves the original position of the binary with the orbital velocity it had at the moment of the explosion, 
that can be as large as $\sim$1,000 km s$^{-1}$ (Geier et al. 2015). The nature of the
companion is poorly understood and it can be a main sequence, subgiant, red giant, AGB, He star or even another white dwarf (Iben
1997; Ruiz-Lapuente 2019). 

The determination of the nature of the progenitor (the exploding star) of type Ia supernovae has been the subject of many studies (e.g. Maoz et al. 2014).
Less effort has gone into understanding the surviving donor companion (e.g. Ruiz-Lapuente 2019).
Only a handful of nearby type Ia supernova remnants (SNRs) have been inspected 
in search of the companion star (Pan et al. 2014).
Importantly, the search for the binary companion of Tycho's SN 1572 has yielded a candidate: a G-type subgiant star labelled Tycho-G (Ruiz-Lapuente 2004; Gonz\'alez-Hern\'andez 2009). The star is relatively close to the center of the supernova remnant, its distance is compatible with it being inside the SNR, it has significantly higher radial velocity and proper motions than 
other stars at the same location in the Galaxy, and it shows signs of pollution from the supernova ejecta. However, other studies have
questioned Tycho-G as the companion donor star because they do not find the expected large
stellar rotation (Kerzendorf et al. 2009), they do not confirm the unusual chemistry (Kerzendorf et al. 2013), or their proposed explosion center does not overlap with the past stellar position at the time of the explosion (Xue \& Schaefer 2015;
Millard et al. 2022). An additional problem is that
the Gaia DR3 (Gaia collaboration et al. 2016; 2022) distance of Tycho-G (Gaia DR3 431160359413315328 in the Gaia designation) is
$1.93^{+0.45}_{-0.31}$ kpc while the distance to the supernova is 2.8$\pm$0.4 kpc (Kozlova \& Blinnikov 2018). Then, the
$\pm$1-$\sigma$ error bars fall close but do not overlap. More accurate distances could help solve this problem.

The lack of clearly confirmed donor companion stars in type Ia SNRs (Schaefer \& Pagnotta 2012; Gonz{\'a}lez Hern{\'a}ndez
et al. 2012; Di Stefano \& Kilic 2012; Pan et al. 2014; Ruiz-Lapuente et al. 2018) has led to a questioning of the
favored  'single-degenerate'  path (mass transfer from a normal companion star to a ultimately exploding
white dwarf) in favor of the 'double-degenerate'
scenario, in which the merger of two white dwarfs produces an explosion that leaves no stellar remnant.

The field of Tycho's SNR has been studied in the radio with emphasis in the morphology and proper motions of the 
extended synchrotron shell 
(Dickel et al. 1991; Reynoso et al. 1997; Williams et al. 2016).
In this paper we present a study of the compact (taken to be $\leq 6''$) radio sources in this field to search for a source 
with large proper motions that could
be related to the companion star. 
Stars of low optical luminosity in the main
sequence are very rarely detected in the radio (Kimball et al. 2009). As noted by Gaidos et al. (2000), the radio emission of normal, single solar analogs rapidly declines to undetectable levels after a few hundred million years. 
On the other hand, it is known that rotation is the main agent responsible for the level of magnetic activity in cool stars (G{\"u}del 2002). Large magnetic fields facilitate the production of radio emission via mechanisms such as gyrosynchrotron. The possible relation between rotation and radio emission is supported by the study of McLean et al. (2012) who find that rapidly 
rotating ($v ~sin ~i ~\geq 20~km ~s^{-1})$ 
ultracool dwarfs have a higher radio detection fraction by about a factor of three compared to objects with 
$v ~sin ~i ~\leq 10~km ~s^{-1}$. 
Also, Lim \& White (1995) detected at 3.6 cm three of the four ultrafast ($v ~sin ~i ~\geq 30~km ~s^{-1})$ 
rotators in the Pleiades. The justification of our search is that the companion donor stars are expected
to have large rotation velocities ($v ~\geq 20~km ~s^{-1}$ for a donor mass $\geq0.2~M_\odot$; Kerzendorf et al. 2009) that could enhance their radio
emission. It should be pointed out, however, that there are a variety
of mechanisms that can produce detectable radio emission in stars
(G{\"u}del 2002; Dzib et al. 2013). In section 2 we present the observations, while in section 3 we discuss some individual
sources. Finally, in section 4 we present our conclusions.

\section{Observations}

\subsection{1.50 GHz Observations for Epoch 2014 February 20}
\label{sec:errors}

The main set of data used in this paper corresponds to observations made
with the Karl G. Jansky Very Large Array (VLA) of NRAO\footnote{The National 
Radio Astronomy Observatory is a facility of the National Science Foundation operated
under cooperative agreement by Associated Universities, Inc.} in the A configuration on 2014 February 20
(2014.140). The observations were taken under project 13A-426 in L-band (0.96-1.97 GHz) using 16 spectral windows, each 64 MHz wide. The amplitude calibrator was J0137+3309 (3C48) and the gain calibrator was J0217+7349. The phase center of the observations was at $RA(J2000) = 00^h~25^m~18\rlap.^{s}966$;
$DEC(J2000) = 64^\circ~08'~20\rlap.{''}41$. 
The data were edited and calibrated following
the standard procedures inside the CASA (Common Astronomy Software Applications;  McMullin et al. 2007) package of NRAO and
the pipeline provided for VLA\footnote{https://science.nrao.edu/facilities/vla/data-processing/pipeline} observations. 

We made images using a robust weighting (Briggs 1995) of 0 and visibilities with baselines larger than 6 k$\lambda$, to suppress structures larger than $\sim$30 arcsec. 
The final image had an rms noise of 28
$\mu$Jy beam$^{-1}$ at the phase center and a synthesized beam of $1\rlap.{''}44\times0\rlap.{''}83$; PA = $-88^\circ$. 
The images were corrected for the primary beam response. The images were also corrected for wide-field effects
using the gridding option  \it widefield \rm with 10$\times$10 subregions in the task TCLEAN. A region with radius of 13 arcmin was inspected for sources with peak emission above 6-$\sigma$ (168 $\mu$Jy beam$^{-1}$). We detected a total of 36 sources. Their positions, flux densities, spectral indices, angular sizes and counterparts, when found, are given in Table 1.
The positions and flux densities were determined using the task IMFIT of
CASA. The spectral indices were determined by splitting the total bandwidth in four bandwidths of 256 MHz each, making
images and fitting the four flux densities obtained for each source as a function of frequency 
with a linear regression in the log-log plane.

\begin{table*}[!t]\centering
  \scriptsize
  \newcommand{\DS}{\hspace{6\tabcolsep}} 
  \begin{changemargin}{-2cm}{-2cm}
    \caption{Parameters of Compact Radio Sources at 1.50 GHz (2014 February 20)} \label{tab:ioniz_av}
    \setlength{\tabnotewidth}{1.0\linewidth}
    \setlength{\tabcolsep}{1.1\tabcolsep} \tablecols{7}
    \begin{tabular}{lcccccc}
      \toprule
  & \multicolumn{2}{c}{Position\tabnotemark{a}}   & Flux & Spectral & Angular Dimensions\tabnotemark{b} &  \\
 No. & RA(J2000)  & DEC(J2000)  & Density($\mu$Jy) & Index & ($\theta_{maj} \times \theta_{min}; PA$) & Counterpart \\
      \cmidrule(r){1-7}
 1 & 00~23~16.800$\pm$0.007  &64~20~25.46$\pm$0.03  & 1808$\pm$200  &  +0.00$\pm$0.52 &  $1.4\pm0.2\times0.9\pm0.3; 31\pm25$ & \nodata  \\     
 2 &  00~23~24.220$\pm$0.010 & 63~59~07.43$\pm$0.02 &  840$\pm$95 &  --1.35$\pm$0.45 &  $\leq0.5$ & \nodata \\
 3 &  00~23~30.774$\pm$0.013 & 64~08~04.34$\pm$0.05  & 528$\pm$75  &  --1.14$\pm$0.48 &  $\leq0.5$ & \nodata \\ 
 4 &  00~23~40.826$\pm$0.011 & 64~10~52.87$\pm$0.03 &  416$\pm$46  &  --1.45$\pm$0.47 &  $\leq0.5$ & \nodata \\
 5 &  00~24~11.264$\pm$0.007 & 64~09~50.02$\pm$0.01 &  521$\pm$31 &  --0.04$\pm$0.30  &  $\leq0.5$ & \nodata \\
 6 & 00~24~18.177$\pm$0.012  & 63~58~58.03$\pm$0.12 & 429$\pm$62  &  +0.02$\pm$0.54 &  $\leq0.5$  & \nodata \\  
 7 & 00~24~30.749$\pm$0.009  & 64~02~00.33$\pm$0.03 & 348$\pm$44  &  --0.59$\pm$0.54 &  $\leq0.5$ & \nodata \\
 8 &  00~24~39.839$\pm$0.013 & 64~01~05.93$\pm$0.05 &  381$\pm$51 &  --1.31$\pm$0.53 &  $\leq0.5$ & \nodata \\
 9 & 00~24~41.071$\pm$0.006 & 64~03~56.91$\pm$0.01 &  596$\pm$37 &  +0.03$\pm$0.30 &  $\leq0.5$  & \nodata \\ 
  10 &  00~24~44.695$\pm$0.006 & 64~04~53.40$\pm$0.05 &  561$\pm$51 &  --1.63$\pm$0.42 & $ \leq0.5$  & \nodata \\
  11 & 00~24~46.084$\pm$0.006  & 64~12~36.46$\pm$0.02 &  528$\pm$37 &  --1.27$\pm$0.34 &  $\leq0.5$ & \nodata \\
  12 & 00~24~51.749$\pm$0.006  & 63~59~08.33$\pm$0.02 &  601$\pm$55 &  --0.89$\pm$0.42 &  $\leq0.5$ & \nodata \\
  13 & 00~24~56.742$\pm$0.007  & 64~17~50.66$\pm$0.03 & 607$\pm$44  &  --1.69$\pm$0.34 &  $\leq0.5$ & \nodata \\
 14 &  00~24~58.631$\pm$0.009 & 64~10~56.82$\pm$0.03 & 486$\pm$53  &  --1.03$\pm$0.37 &  $\leq0.5$ & \nodata \\
  15 &  00~25~01.637$\pm$0.005 & 64~07 47.75$\pm$0.01 &  843$\pm$46 &  --1.49$\pm$0.20 &  $\leq0.5$ & \nodata \\ 
  16 & 00~25~10.166$\pm$0.003 & 64~16~28.06$\pm$0.01 & 1258$\pm$53  &  --0.64$\pm$0.20 & $\leq0.5$ & \nodata \\ 
 17 & 00~25~14.131$\pm$0.027  & 64~00~15.08$\pm$0.12 &  1652$\pm$288 &  --1.98$\pm$0.61 &  $4.8\pm1.1\times2.2\pm0.7; 45\pm14$ & \nodata \\
 18 &  00~25~16.087$\pm$0.016 & 63~59~40.96$\pm$0.14 &  796$\pm$108 &  --0.99$\pm$0.63 &  $2.3\pm0.5\times1.3\pm0.5; 168\pm27$ & \nodata \\
 19 & 00~25~19.270$\pm$0.001  &  64~08~53.28$\pm$0.01& 5150$\pm$68  &  --1.30$\pm$0.06 &  $\leq0.5$ & [RMG97] 002231.23+635216.9 \\  
 20 &  00~25~26.037$\pm$0.002 & 64~04~40.89$\pm$0.01  & 3575$\pm$90  &  --1.11$\pm$0.08 & $\leq0.5$  & [RMG97] 002237.96+634804.7 \\
 21 &  00~25~34.981$\pm$0.004 & 64~08~43.06$\pm$0.01 &  741$\pm$33 &  --0.70$\pm$0.27 & $\leq0.5$ & \nodata \\
 22 & 00~25~36.024$\pm$0.007  & 64~10~39.81$\pm$0.01 &  330$\pm$33 &  --1.31$\pm$0.52 &  $\leq0.5$ & \nodata \\
 23 &  00~25~40.204$\pm$0.005 & 64~13~05.95$\pm$0.01 & 537$\pm$40  &  --0.10$\pm$0.35 & $\leq0.5$   & \nodata \\
 24 & 00~25~40.601$\pm$0.006  & 64~11~33.72$\pm$0.02 & 607$\pm$53  &  --0.92$\pm$0.33 &  $\leq0.5$ & \nodata \\
 25 & 00~25~50.714$\pm$0.022  & 64~09~28.82$\pm$0.17 & 1058$\pm$115 & --0.95$\pm$0.49 & $4.6\pm0.7\times2.3\pm0.5; 
 7\pm11$ & Nucleus of radio galaxy \\
 26 &  00~26~05.112$\pm$0.002 & 64~17~13.89$\pm$0.01 &  2306$\pm$81 & --0.93$\pm$0.14  & $\leq0.5$  & \nodata \\
 27 &  00~26~08.967$\pm$0.006 & 64~16~33.89$\pm$0.02 &  590$\pm$51 & --0.98$\pm$0.36  &  $\leq0.5$ & \nodata \\
 28 & 00~26~16.279$\pm$0.014  &  64~21~33.46$\pm$0.14 &  3476$\pm$209 & --1.60$\pm$0.40  &  $5.7\pm0.6\times1.2\pm0.3; 29\pm3$  & Disk galaxy 2MFGC 305  \\
 29 &  00~26~23.102$\pm$0.007 & 64~12~32.06$\pm$0.01 &  581$\pm$35 & --1.61$\pm$0.35 &  $\leq0.5$ & \nodata \\
 30 & 00~26~23.656$\pm$0.010  & 64~04~42.34$\pm$0.04 &  253$\pm$46 &  --1.33$\pm$0.55 &  $\leq0.5$ & \nodata \\
 31 & 00~26~24.059$\pm$0.003  & 63~56~24.03$\pm$0.01 &  1725$\pm$123 &  --0.30$\pm$0.32 &  $1.0\pm0.1\times0.6\pm0.3;  0\pm17$ &  \nodata \\
 32 &  00~26~28.133$\pm$0.012 & 63~58~36.78$\pm$0.08 &  508$\pm$62 &  +1.05$\pm$0.39 &  $\leq0.5$  & \nodata \\
 33 &  00~26~30.651$\pm$0.017 & 64~14~36.13$\pm$0.06 & 352$\pm$57  & +0.49$\pm$0.62  &  $\leq0.5$ & \nodata \\
 34 &  00~26~50.729$\pm$0.002 & 64~10~25.52$\pm$0.01 & 3324$\pm$112  &  --1.01$\pm$0.14 &  $1.0\pm0.1\times0.6\pm0.1;  62\pm7$  & \nodata  \\
 35 &  00~26~51.389$\pm$0.002 & 64~02~20.16$\pm$0.01 &  20284$\pm$682 &  --0.99$\pm$0.11 &  $1.3\pm0.1\times0.8\pm0.1; 69\pm1$ & Radio source EQ 0024+638 \\
 36 &   00~27~11.599$\pm$0.017 & 64~12~44.78$\pm$0.02 &  462$\pm$73 &  +0.25$\pm$0.55 &  $\leq0.5$ & \nodata \\
           \bottomrule
      \tabnotetext{a}{\small Right ascension (RA) is given in hours, minutes, and seconds. 
      Declination (DEC) is given in degrees, arcminutes, and arcseconds.}
      \tabnotetext{b}{\small The deconvolved major ($\theta_{maj}$) and minor ($\theta_{min}$) axes are given in arcsec. The position angle ($PA$) is given in degrees.}
    \end{tabular}
  \end{changemargin}
\end{table*}

Following the procedure of Anglada et al. (1998) we determine that 37$\pm$6 background sources were expected. This estimate suggests that most, and probably all, of the sources detected are background objects 
unrelated to the remnant region.
Nevertheless, within the uncertainty, it cannot be ruled out that a few sources could be associated with the
remnant.

\subsection{Observations with the Classic VLA}
\label{sec:command}

To gain information on the time variability and possible proper motions of the sources found
in the 2014 image, we analyzed classic VLA observations made in the A configuration on the three epochs listed in Table 2. 
As in the 2014 data, the amplitude calibrator was J0137+3309 (3C48) and the gain calibrator was J0217+7349
for the three epochs. The data was concatenated to gain signal to noise ratio and assigned an average epoch of 1993.468 and
an average frequency of 1.43 GHz. The data were edited and calibrated using the software package
Astronomical Image Processing System (AIPS) of NRAO.
We made images using a robust weighting (Briggs 1995) of 0 and visibilities with baselines larger than 6 k$\lambda$, to suppress structures larger than $\sim$30 arcsec. 

\begin{table*}[!t]\centering
  \small
  \newcommand{\DS}{\hspace{4\tabcolsep}} 
  \begin{changemargin}{-2cm}{-1cm}
    \caption{Observations with the Classic VLA in A Configuration} \label{tab:sources2}
    \setlength{\tabnotewidth}{1.0\linewidth}
    \setlength{\tabcolsep}{1.0\tabcolsep} \tablecols{6}
    \begin{tabular}{lccccc}
      \toprule
    &  & Amplitude  & Gain & Frequency & Bandwidth \\
 Project  & Epoch & Calibrator & Calibrator & (GHz) & (MHz)\\
 \cmidrule(r){1-6}
  AV84 &  1983 Nov 13 (1983.868) & J0137+3309  &  J0217+7349  & 1.38  & 6.25 \\
  AM437 &  1994 Mar 18 (1994.211) & J0137+3309   & J0217+7349   & 1.51 & 6.25 \\
  AR464 &  2002 Apr 29 (2002.326) & J0137+3309   &  J0217+7349  & 1.41  & 6.25 \\
           \bottomrule
    \end{tabular}
  \end{changemargin}
\end{table*}

The final image had an rms noise of 75
$\mu$Jy beam$^{-1}$ at the phase center and a synthesized beam
of $1\rlap.{''}39\times1\rlap.{''}33$; PA = $+28^\circ$. As for the 2014 image,
a region with radius of 13 arcmin was inspected for sources with peak emission above 6-$\sigma$ (450 $\mu$Jy beam$^{-1}$). We detected a total of 17 sources.  Again, following the procedure of Anglada et al. (1998) we determine that 18$\pm$4 background sources were expected. As in the more sensitive 2014 data, this estimate suggests that most, and probably all, of the sources detected are background objects unrelated to the remnant region.
Their positions and flux densities are given in Table 3. In this Table we also list the difference in
RA and DEC between the 2014 positions and those obtained from the classic VLA data.

\begin{table*}[!t]\centering
   \small
  \newcommand{\DS}{\hspace{4\tabcolsep}} 
  \begin{changemargin}{-2cm}{-1cm}
    \caption{Parameters of Compact Radio Sources from Classic VLA Data} \label{tab:sources2}
    \setlength{\tabnotewidth}{1.0\linewidth}
    \setlength{\tabcolsep}{1.0\tabcolsep} \tablecols{7}
    \begin{tabular}{lcccccc}
      \toprule
  & \multicolumn{2}{c}{Position\tabnotemark{a}}   & Flux & $\Delta$RA\tabnotemark{b} & $\Delta$DEC\tabnotemark{b} & Flux Ratio\\
 No. & RA(J2000)  & DEC(J2000)  & Density($\mu$Jy) & (seconds) & (arcseconds) & (2014/1983) \\
      \cmidrule(r){1-7}
  6 &  00  24  18.052$\pm$0.099 & 63  58  58.73$\pm$1.02 &
  1903$\pm$414 &  0.125$\pm$0.099 & -0.70$\pm$1.03 &
  0.23$\pm$0.06 \\
 9 &  00  24  41.071$\pm$0.014 & 64  03 56.99$\pm$0.09 &
  663$\pm$163 &  0.000$\pm$0.015 & -0.08$\pm$0.09 &
  0.90$\pm$0.23 \\
 10 &  00 24  44.701$\pm$0.013 & 64 04  53.39$\pm$0.15 &
  692$\pm$182 &  -0.006$\pm$ 0.010 & 0.01$\pm$0.16 &
  0.87$\pm$0.24 \\
 12 &  00  24  51.749$\pm$0.016 &  63  59  07.99$\pm$0.17 &
  815$\pm$107 &  0.000$\pm$0.017 &  0.34$\pm$0.17 &
  0.77$\pm$0.12 \\
 13 &  00  24  56.718$\pm$0.023 &  64 17 50.48$\pm$0.16 &  582$\pm$193 
  &  0.024$\pm$0.024 &  0.18$\pm$0.16 &  1.13$\pm$0.38 \\
 15 &  00  25  01.648$\pm$0.009 &  64 07  47.89$\pm$0.07 &
  1130$\pm$179 & -0.011$\pm$0.010 & -0.14$\pm$0.07 &
  0.80 $\pm$0.13 \\
 16 &  00  25  10.161$\pm$0.008 &  64 16 27.85$\pm$0.06
  &  1735$\pm$197 &  0.005$\pm$0.009 &  0.21$\pm$0.06&
  0.75$\pm$0.09 \\
 17 &  00  25  14.076$\pm$0.097 &  64 00  13.84$\pm$0.69 &
  3782$\pm$530 &  0.055$\pm$ 0.101 &  1.24$\pm$0.70&
  0.48$\pm$0.11 \\
 19 &  00  25  19.282$\pm$0.001 &  64 08 53.40$\pm$0.01
  &  6657$\pm$166 & -0.012$\pm$0.001 & -0.12$\pm$0.01 &
  0.82$\pm$0.02 \\
 20 &  00  25  26.046$\pm$0.002 &  64 04 40.97$\pm$0.02
  &  3831$\pm$168 &  -0.010$\pm$0.003 & -0.08$\pm$0.02 &
  0.98  $\pm$0.05 \\
 21 &  00  25  34.987$\pm$0.011 &  64 08 43.23$\pm$0.07 &
  974$\pm$173 &  -0.006$\pm$0.012 & -0.17$\pm$ 0.07 &  0.79$\pm$0.14 \\
 24 &  00 25  40.605$\pm$0.017 &  64 11 34.00$\pm$0.11 &  675$\pm$176
  &  -0.004$\pm$0.018 & -0.28$\pm$0.11 &  0.94$\pm$ 0.26 \\
 26 &  00  26  05.110$\pm$0.009 &  64 17 13.62$\pm$0.05
  &  2229$\pm$216 &  0.002$\pm$0.009 &  0.27$\pm$0.06 &
  1.08$\pm$0.11 \\
 28 & 00 26  16.278$\pm$0.053 &  64 21 32.73$\pm$0.35 &
  6157$\pm$1010 &  0.012$\pm$0.055 &  0.57$\pm$0.37 &
  0.61$\pm$0.11 \\
 31 &  00  26  24.054$\pm$0.014 &  63 56 23.45$\pm$0.07
  &  1338$\pm$214 &  0.005$\pm$ 0.014 &  0.58$\pm$ 0.07 &
  1.31$\pm$0.23 \\
 34 &  00 26  50.724$\pm$0.005 & 64 10 25.28$\pm$0.03
  &  3564$\pm$222 &  0.005$\pm$0.005 &  0.24$\pm$0.03 &
  0.98$\pm$0.08 \\
 35 &  00  26  51.364$\pm$0.002 &  64 02 19.78$\pm$0.01
  &  20092$\pm$263 &  0.025$\pm$0.003 &  0.38$\pm$0.01 &
  1.06$\pm$0.04 \\
           \bottomrule
      \tabnotetext{a}{\small Right ascension (RA) is given in hours, minutes, and seconds. 
      Declination (DEC) is given in degrees, arcminutes, and arcseconds.}
        \tabnotetext{b}{\small Position differences given by 2014 positions minus classic VLA positions.}
    \end{tabular}
  \end{changemargin}
\end{table*}

All 17 sources detected in the image from the classic VLA data 
were detected in the more sensitive (by a factor of 2.7) 2014 observations. To compare the flux densities
at the two epochs we corrected the 2014  flux densities from 1.50 to 1.43 GHz using the spectral indices listed in Table 1.
This correction is small, of order 5-10\%.
In the last column of Table 3 we list the ratio of the flux densities. Eight of the 17 sources have a ratio consistent
within noise with 1 and thus no
variability. Only source 6 shows a variability larger than a factor of 2, decreasing by a factor of 4 between databases.


\subsection{Upper Limits to the Proper Motions}
\label{sec:UpperLimits}

Analyzing Table 3 we find that the position differences between the 2014 image and the classic VLA image (columns 
5 and 6) are above a conservative 5-$\sigma$ value only for sources 19, 31, 34 and 35. Should these possible
small proper motions be considered
as real? Sources 31, 34 and 35 are resolved angularly (see Table 1) and this characteristic could well account for the small
difference in position between epochs due to differences in the beam shape. 
The position differences for the remaining source 19 are of order of only $\sim0\rlap.{''}1$.  
Taking this displacement as an upper limit to the true proper motions, we can crudely set an upper limit
to the velocity of the sources in the plane of the sky. 
Given the time interval of 20.672 yr between the two datasets and a distance of 2.8 kpc to the supernova 
(Kozlova \& Blinnikov 2018) this results in an upper limit of 65 km s$^{-1}$ for the velocity of the radio sources in the plane of the sky,
if they were located at the distance of the supernova. As noted before, it is most probable that all sources are extragalactic
and that the small position differences measured are the result of noise and not true displacements. Supporting the extragalactic interpretation is that none of the 36 radio sources has a counterpart in the Gaia DR3 catalog (Gaia collaboration et al. 2016; 2022).
Tycho's SNR is almost in the plane of the Galaxy (it has a galactic latitude of $1\rlap.^\circ4$). Then, extragalactic objects in this
direction will be heavily
obscured in the optical and very difficult to detect in the observations of the Gaia project.

\subsection{The Non Detection of Tycho-B}
\label{sec:NonDetection}

The star Tycho-G was not detected in our observations. Perhaps more interestingly, the star Tycho-B was not detected either.
Tycho-B is a fast-rotator A-type star that has been proposed as a candidate to be the donor companion of the 
exploding white dwarf that produced SN 1572 (Kerzendorf et al. 2013), although this proposal has been questioned later
(Kerzendorf et al. 2018). 

The best studied case of radio emission from a rapidly rotating A-type star is that of Altair ($\alpha$ Aql; White et al. 2021).
From the model of these authors we estimate that Altair is expected to have a flux density of $\simeq 10~\mu Jy$ at the frequency
of 1.5 GHz. Tycho-B, with position $RA(J2000) = 00^h~ 25^m~ 19\rlap.^s985;~ 
DEC(J2000) = +64^\circ~ 08' 17\rlap.{''}19$, is located at a distance of 2.7 kpc (Gaia Collaboration 2020). On the other hand, Altair is
located at a distance of only 5.1 pc (van Leeuwen 2007). We then conclude that if Tycho-B has a similar radio emission to that
of Altair, its flux density would be $(2700/5.1)^2 =  2.8 \times 10^5$ weaker and undetectable with present instrumentation.

\section{Individual Sources}
\label{Sources}

To gain information on the sources detected we used the SIMBAD database to search for counterparts within a few arcsec of the 
radio position.

\subsection{Sources 19 and 20}

These sources are reported by Reynoso et al. (1997) in their study of the expansion of the supernova remnant.
They used them to align their images at different epochs. Their compact size and negative spectral index (see Table 1)
suggest they are background active galactic nuclei.

\subsection{Source 25}

This source seems to be the nucleus of a radio galaxy (see Figure 1). The lobes of the radio galaxy extend over $\sim 35''$
in the north-south direction. Since they are clearly extended, the lobes
are not reported in our list of compact sources.
The nucleus is elongated along the axis defined by the radio lobes and probably traces more recent ejections.

\begin{figure}[!t]
  \includegraphics[width=\columnwidth]{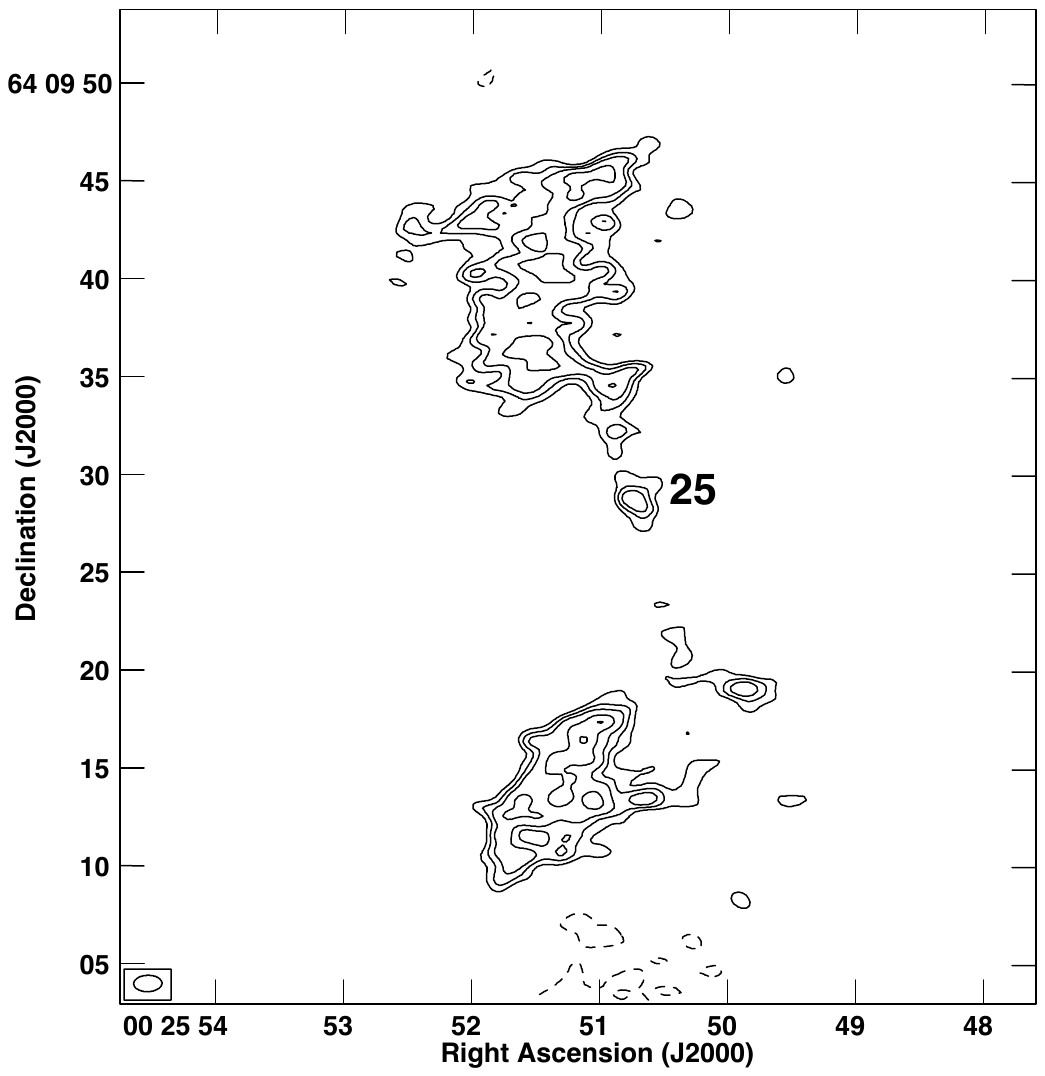}
  \vskip-3.0cm
  \caption{Radio galaxy related to source 25. Contours are -4, 4, 5, 6, 8 and 10 times 17 $ \mu$Jy beam$^{-1}$, the rms noise 
  of this region of the image. The synthesized beam ($1\rlap.{''}44 \times  0\rlap.{''}83; -88^ \circ$) is shown in the bottom left corner.}
  \label{fig:simple}
\end{figure}

\subsection{Source 28}

This source was initially catalogued in the 2MASS-selected Flat Galaxy Catalog (2MFGC; Mitronova et al. 2006).
The sources in this catalog are expected to be disk galaxies 
selected from the Extended Source Catalog of the Two Micron All-Sky Survey (XSC 2MASS; 
Jarrett et al. 2000) on the basis
of their 2MASS axial ratio being a/b$\geq$3. We find that the infrared and radio images approximately
coincide and show similar elongated morphology (Figure 2). 

\begin{figure}[!t]
  \includegraphics[width=\columnwidth]{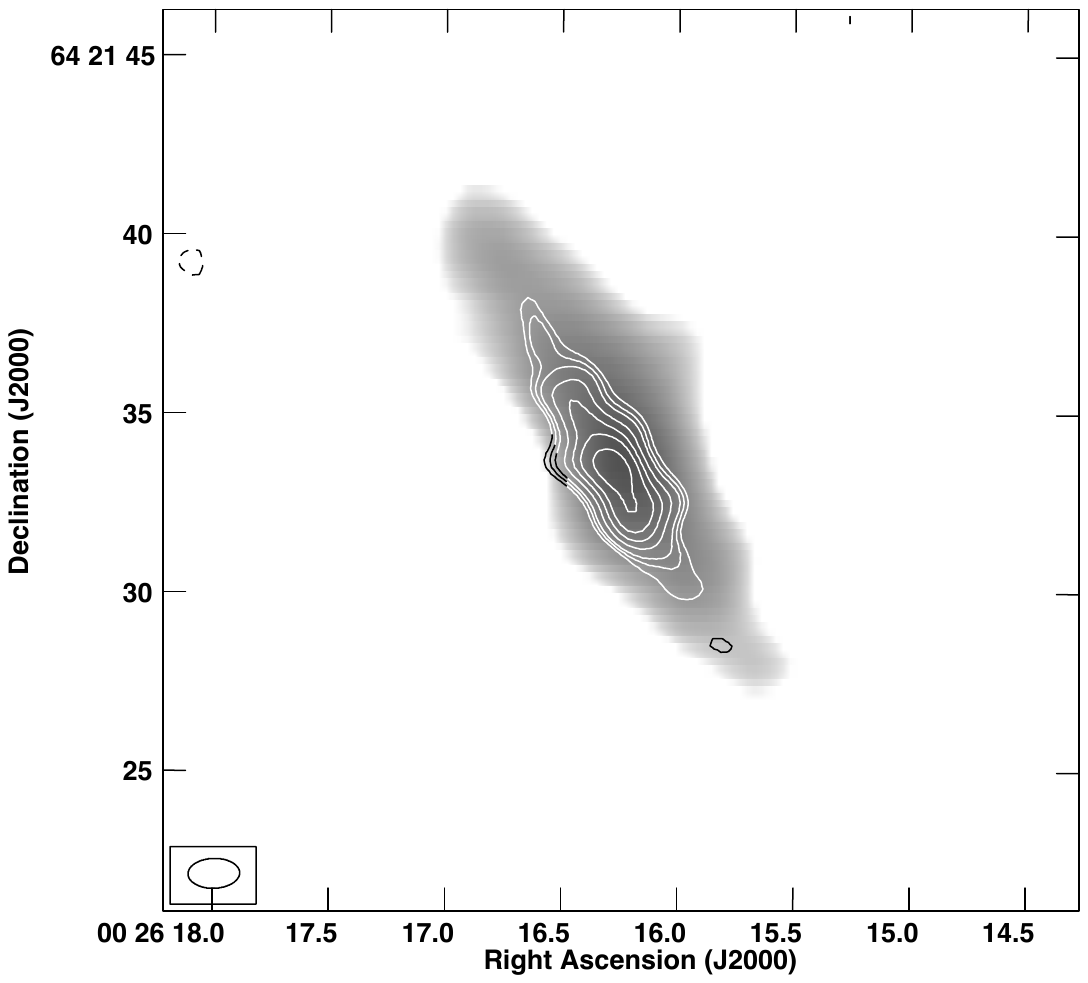}
  \vskip-2.3cm
  \caption{The radio source 28 is shown in white contours in this image. The 
  contours are -4, 4, 5, 6, 8, 10, 12 and 15  times 11 $ \mu$Jy beam$^{-1}$, the rms noise 
  of this region of the image. The synthesized beam ($1\rlap.{''}44 \times  0\rlap.{''}83; -88^ \circ$) is shown in the bottom left corner.
  The infrared image from the Two Micron All Sky Survey (2MASS) in K$_s$ band is shown in greyscale.}
  \label{fig:simple2}
\end{figure}

\subsection{Source 32}

This source is interesting because it shows a clearly positive spectral index (+1.05$\pm$0.39; Table 1). All other sources in Table 1 have spectral indices that are consistent within noise with a negative spectral index. Background radio sources usually
show clearly negative spectral indices, taken to be indicative of optically-thin synchrotron emission. For example,
Smol{\v{c}}i{\'c} et al. (2016) obtained accurate spectral indices for 159 sources by combining 
Australia Telescope Compact Array (ATCA) 2.1 GHz observations with Sydney University Molonglo Sky Survey (SUMSS) 843 MHz data. Only three of the sources show positive spectral indices ($\alpha > 0$). 

Source 32 could be a
high frequency peaker (HFP). These are compact, sometimes powerful
extragalactic radio sources with well-defined peaks in their radio
spectra above a few GHz, with most of them being high redshift
quasars (Dallacasa et al. 2000). A possible explanation for
the HFP radio sources is that we are observing synchrotron emission produced by blazars caught during a flare, when a 
highly self-absorbed (optically-thick) component dominated the emission (Tinti et al. 2005)

\subsection{Source 35}

This is the brightest compact source in the region considered. It was reported previously by Schwarz et al. (1995) from HI 
VLA observations made in 1993 in the C configuration. These authors proposed that it is a background source.
It is an elongated source (see Table 1) and it could be a disk galaxy seen edge-on or an unresolved radio galaxy.



\section{Conclusions}

Our main conclusions are the following:

1) We compared observations made in L-band toward the Tycho supernova remnant in 2014 with the Jansky VLA and
previously with the classic VLA. The purpose of this comparison was to
search for radio sources showing large proper motions that could be related to the donor companion of the exploding white dwarf that produced the supernova in 1572. We failed to find radio sources with large proper motions 
($\geq$ 65 km s$^{-1}$) and conclude that
most, and probably all, of the sources detected are background extragalactic objects. However, we argue that these donor
companions could be radio sources and believe that its search in other type Ia supernova remnants
is worthwhile.

2) We discussed several sources individually and concluded that their characteristics are
compatible with them being most likely background extragalactic sources. Source 28 is most probably a disk galaxy seen edge-on. 
Source 32 could be a high frequency peaker, possibly a blazar caught during a flare.

\section{Acknowledgments}

This publication makes use of data products from the Two Micron All Sky Survey, which is a joint project of the University of Massachusetts and the Infrared Processing and Analysis Center, funded by the National Aeronautics and Space Administration and the National Science Foundation. This research has made use of the SIMBAD database, operated at CDS, Strasbourg, France.

\end{document}